\documentclass[preprint,prd, aps,tighten,nofootinbib,amssymb]{revtex4}

\usepackage{epsf,epsfig,subfigure,graphicx,multirow,amsmath,amssymb,amsfonts,amssymb}
\usepackage{graphicx,color}
\usepackage{hyperref}
\usepackage{cancel}
\usepackage{epstopdf}

\def\beq{\begin{equation}}
\def\eeq{\end{equation}}
\def\bea{\begin{eqnarray}}
\def\eea{\end{eqnarray}}
\def\ba{\begin{array}}
\def\ea{\end{array}}
\def\bec{\begin{center}}
\def\ec{\end{center}}

\newcommand{\be}{\begin{equation}}
\newcommand{\ee}{\end{equation}}

\newcommand{\dis}[1]{\begin{equation}\begin{split}#1\end{split}\end{equation}}

\newcommand{\gev}{\,\textrm{GeV}}

\begin{document}


\preprint{CTPU-15-05}
\title{\Large\bf  Communication with SIMP dark mesons via $Z'$-portal }

\author{Hyun Min Lee$^{a}$\footnote{e-mail: hminlee@cau.ac.kr} and 
Min-Seok Seo$^{b}$\footnote{e-mail: minseokseo@ibs.re.kr}
 }
\affiliation{
$^{a}$\it Department of Physics, Chung-Ang University, Seoul 156-756, Korea \\
$^{b}$\it Center for Theoretical Physics of the Universe, Institute for Basic Science (IBS), Daejeon 305-811, Korea
\vskip 1.0cm
}




\begin{abstract}
 
 We consider a consistent extension of the SIMP models with dark mesons by including a dark U(1)$_D$ gauge symmetry.   
Dark matter density is determined by a thermal freeze-out of the $3\to2$ self-annihilation process, thanks to the Wess-Zumino-Witten term.  
 In the presence of a gauge kinetic mixing between the dark photon and the SM hypercharge gauge boson,  dark mesons can undergo a sufficient scattering off the Standard Model particles and keep in kinetic equilibrium until freeze-out in this SIMP scenario. 
Taking the SU$(N_f)\times$SU$(N_f)/$SU$(N_f)$ flavor symmetry under the SU($N_c$) confining group, we show how much complementary the SIMP constraints on the parameters of the dark photon are  for current experimental searches for dark photon.    
  

\end{abstract}


\maketitle




\section{Introduction}\label{Sec:Introduction}

Various evidences of dark matter (DM) imply that fundamental particles and interactions need to be extended beyond the Standard Model (SM) \cite{Bertone:2004pz}.
 One of the appealing suggestions is the thermal DM scenario, where the DM relic density is determined through the freeze-out of the DM number changing process.
Weakly Interacting Massive Particle (WIMP) provides the most popular thermal DM scenario, in which the annihilation of a DM pair into a pair of SM particles is responsible for the freeze-out.
Since WIMP mass is of order weak scale for the effective coupling of $\alpha_{\rm eff} \sim {\cal O}(10^{-2})$, `WIMP miracle' has been the mainstream for thermal DM studies, corroborating the expectation of finding new physics at the weak scale in the solutions for gauge hierarchy problem.
 
Another interesting proposal for thermal (pseudo-) scalar DM has been recently made under the name of Strongly Interacting Massive Particle (SIMP) \cite{Hochberg:2014dra,Hochberg:2014kqa,Bernal:2015bla}, explaining DM relic density through the freeze-out of $3\to 2$ self-annihilation.
The DM self-interaction is motivated by potential small-scale problems \cite{Spergel:1999mh}, although it is strongly constrained by bullet cluster \cite{Markevitch:2003at} and simulations on halo shape \cite{Rocha:2012jg}.
  As a result, the SIMP scenario predicts dark matter with dimensionless self-interacting coupling of order one and mass in the $0.1 - 1$ GeV range, which has not been explored seriously so far. 
  
The SIMP scenario requires the interaction between dark and SM sectors in the form of scattering for dark matter to be in kinetic equilibrium with the SM particles, without altering the structure formation \cite{Carlson:1992fn}.
 Since such an inter-sector interaction also leads to the DM annihilation into SM particles, the inter-sector interaction strength is bounded from above for the dominance of $3\to 2$ self-annihilation, if combined, resulting in $ n_{\rm DM}\langle \sigma v \rangle_{\rm ann} < n_{\rm DM}^2\langle \sigma v^2 \rangle_{3\to2} <n_{\rm SM}\langle \sigma v \rangle_{\rm scatt}$, at the freeze-out temperature. 
Taking other constraints from ground-based experiments into account in addition, we can make quite a concrete prediction on the parameters of a specific SIMP model.

  In this article, we consider a SIMP model with dark mesons suggested 
  in Ref.~\cite{Hochberg:2014kqa},  where the 5-point interactions between dark mesons for $3\to2$ annihilation come from the leading interactions of the Wess-Zumino-Witten (WZW) term \cite{Wess:1971yu, Witten:1983tw}.
 From the model point of view,  the WZW term is interesting because it encodes various aspects on dark sector, namely, the WZW term exists only for a specific flavor symmetry of light dark quarks, depending on its spontaneous breaking pattern \cite{Witten:1983tx}, and it contains color number as a topological index \cite{Witten:1983tw}.
 As an inter-sector interaction, we consider the hidden valley scenario \cite{Strassler:2006im}, that some heavy dark sector particle has a renormalizable coupling to the mediator particle that communicates between DM and SM particles.
Higgs-portal interaction would be a natural candidate, but it is not enough for a sufficiently large DM-SM particle scattering at freeze-out temperature due to small Yukawa couplings of the light SM fermions. Thus, we study the case with a gauge kinetic mixing \cite{Okun:1982xi}, that is the renormalizable  and gauge invariant interaction between SM hypercharge U(1)$_Y$ and dark sector U(1)$_D$ with a dimensionless coupling.
When the dark U(1) gauge coupling is not too tiny,  dark meson annihilation into dark photons could easily dominate the annihilation process of dark matter. But, we can forbid it by taking the dark photon to be heavier than dark mesons. In this case,  the gauge kinetic mixing plays a role of `hidden valley'  in the SIMP scenario.
  
 In Sec. \ref{Sec:SIMPreview}, we briefly review dark mesons in the SIMP scenario, the abundance of which is frozen out by the WZW term.
 In Sec. \ref{Sec:darkU(1)}, we discuss properties of the dark U(1) gauge symmetry, U(1)$_D$, that is compatible with both the WZW term and the SIMP scenario.
 By taking SU($N_c$) as an example of confining gauge group,  in Sec. \ref{Sec:kineticmix}, we present a viable parameter range for the dark photon mass and the strength of the gauge kinetic mixing. 
 Finally, conclusions are drawn.

\section{SIMP dark mesons and the WZW term }\label{Sec:SIMPreview}

Dark meson appears as a composite state of dark quarks in models with dark confining gauge group $G_c$ and it has several interesting properties \cite{Cline:2013zca}.
 The lightest mesons are interpreted as pseudo-Goldstone bosons from a spontaneous breaking of (accidental) flavor symmetry, which guarantees their stability.
 Whereas flavor symmetry is broken by higher dimensional operators, due to the compositeness of dark mesons, their decays induced by higher dimensional operator is suppressed, as compared to those of a fundamental scalar, so the stability of dark mesons is easier to achieve.
For instance, dark mesons can be unstable by the decay into lepton pairs $\pi_a \to 
 \ell \overline{\ell}$, due to the dimension-6 four-Fermi interaction,
 \dis{\frac{1}{M^2}(\overline{q}\gamma^5 \gamma_\mu T^a q)(\overline{\ell}\gamma^\mu \ell)\sim \frac{F}{M^2}\partial_\mu \pi^a (\overline{\ell}\gamma^\mu \ell),}
  where $M$ is the scale at which an explicit breaking of flavor symmetry occurs, $T^a$ is the generator of flavor symmetry, and $F$ is the dark meson decay constant of order of the confining scale $\Lambda$.
  Note that this operator is dimension-6, rather than dimension-5, which is the dimension of the corresponding operator for a fundamental pseudo-scalar DM. 
  Then, the lifetime of dark mesons is given by $\Gamma^{-1}\simeq 8\pi M^4/(F^2 m_\pi m_\ell^2)$, and, for several hundred MeV to GeV-scale dark mesons and a similar confining scale, $\pi_a \to \mu \overline{\mu}$ provides the largest decay rate.
   It is longer than the age of universe, $6.6\times 10^{41}$GeV$^{-1}$, as long as $M$ is larger than $10^9$ GeV. Dark mesons can also decay into a pair of photons due to a dimension-7 operator, $\frac{1}{M^3}(\overline{q}\gamma^5 T^a q) F_{\mu\nu}{\tilde F}^{\mu\nu}\sim \frac{\Lambda^3}{M^3 F}\pi^aF_{\mu\nu}{\tilde F}^{\mu\nu}$, but the lifetime estimated is $\Gamma^{-1}\simeq \pi F^2M^6/(m_\pi^3\Lambda^6)$, resulting in a less stringent constraint on the cutoff, $M>10^7$ GeV.
 Moreover, the interactions between dark mesons induce the DM self-scattering, which provides a solution to the small-scale problems such as `core-cusp' or `too-big-to-fail' problems.
 
 In the SIMP scenario proposed in Ref. \cite{Hochberg:2014dra}, the DM relic density can be explained from the freeze-out of dark mesons in the presence of their 5-point self-interactions, provided by the WZW term \cite{Hochberg:2014kqa},
 \begin{widetext}
 \dis{S_{\rm WZW}&=-\frac{i N_c}{240\pi^2}\int d \Sigma^{ijklm} {\rm Tr} [ U^{-1}\partial_i U U^{-1} \partial_j U
 U^{-1} \partial_k U U^{-1} \partial_l U U^{-1} \partial_m U]
\\
&=\frac{N_c}{240\pi^2}\int d^4x \epsilon^{\mu\nu\rho\sigma} \pi^a \partial_\mu \pi^b \partial_\nu \pi^c \partial_\rho \pi^d \partial_\sigma \pi^e
{\rm Tr}(T_aT_bT_cT_dT_e)
 +\cdots.}  
 \end{widetext}
 The WZW term is the outcome of a specific flavor symmetry $G_f$ and its spontaneous breaking to the subgroup $H$ for a given confining gauge group, relying on a nontrivial fifth homotopy group, $\pi_5(G_f/H)={\mathbb Z}$ \cite{Witten:1983tx}.
 The manifest non-chiral global symmetry $H$ is unbroken if it is respected by dark quark masses \cite{Vafa:1983tf}. As a consequence, degenerate dark quark masses $m_q$, or degenerate dark meson mass $m_\pi^2=8(\Lambda^3/F^2)m_q$ guarantee the existence of the WZW term. 
 
 Here, we quote the results obtained in Ref. \cite{Hochberg:2014kqa}, which will be useful for the discussion hereafter.
 The $3\to2$ annihilation cross section is calculated from the WZW term to be
 \dis{\langle \sigma v^2 \rangle_{3\to 2}=\frac{5\sqrt{5}N_c^2 m_{\pi}^5}{2\pi^5 F^{10}}\frac{t^2}{N_\pi^3} \left(\frac{T_F}{m_\pi}\right)^2,}
 where $T_F$ is the freeze-out temperature, $N_\pi$ is the number of dark mesons, or dim$(G_f/H)$, and $t^2$ is a factor determined by group theory, $\sim N_f^5$ for large $N_f$.
 As a result of freeze-out, the number density of dark matter is given by
 $n_{\rm DM}=c T_{\rm eq}s / m_{\pi}$,
 where $T_{\rm eq}=0.8$ eV is the matter-radiation equality temperature, $s=(2\pi^2/45)g_{*S}(T)T^3$ is the entropy density of relativistic particles in thermal equilibrium, and $c\simeq 0.54$ is the numerical constant. For $\langle \sigma v^2 \rangle_{3\to 2}\equiv {\alpha^3_{\rm eff}}/{m^5_{\rm DM}}$,  the freeze-out condition for the $3\to2$ annihilation,
  $n_{\rm DM}^2(T_F)\langle \sigma v^2 \rangle_{3\to 2}(T_F)=H(T_F)$, with the freeze-out temperature at $T_F \simeq m_\pi/20$ \cite{Hochberg:2014dra},  determines dark matter mass in terms of the effective DM self-coupling.
 \be
 m_{\rm DM} \simeq 0.03\, \alpha_{\rm eff} (T^2_{\rm eq} M_P)^{1/3}. 
 \ee 
 Thus, for $\alpha_{\rm eff}=1-10$, we get $m_{\rm DM}=35-350\,{\rm MeV}$. 
As will be discussed in a later section, in order to keep dark matter in kinetic equilibrium with heat bath, it is necessary to introduce the inter-sector interaction between dark and SM sectors. 
 
 On the other hand, the leading $2\to2$ self-scattering comes from the kinetic term $(F^2/16){\rm Tr}(\partial_\mu U \partial^\mu U^{-1})$, whose cross section is given by
 \dis{ \sigma_{\rm self}=\frac{m_\pi^2}{32\pi F^4}\frac{a^2}{N_\pi^2},}
 where $a^2$ is another group theory factor $\sim N_f^4$ for large $N_f$.
 The self-interaction cross section is constrained to be $\sigma_{\rm self}/m_\pi \lesssim 1 {\rm cm}^2/{\rm g}$.
  This condition, together with the perturbativity bound of chiral perturbation theory, $m_\pi/F<2\pi$, imposes the dark meson mass to be in the $0.1 - 1$ GeV range, depending on the confining gauge group. 
 The group theory factors for possible gauge and flavor symmetries with nonzero WZW terms are summarized in Table \ref{Table:WZWsummary}. 
 
 \begin{widetext}
\begin{table}
\begin{center}
\begin{tabular}{| c || c | c | c | c |} 
\hline
  \scriptsize  $G_c$ & \scriptsize $G_f/H$ & \scriptsize  $N_\pi$ & \scriptsize $t^2$ & \scriptsize $N_f^2 a^2$ \\ \hline \hline
\scriptsize SU($N_c$) & \scriptsize $ \begin{array}{l}
\frac{{\rm SU}(N_f)\times{\rm SU}(N_f)}{{\rm SU}(N_f)}\\
\quad\quad {\tiny (N_f\ge 3)} 
\end{array}$
 & \scriptsize $N_f^2-1$ & \scriptsize $\frac43N_f(N_f^2-1)(N_f^2-4)$ & \scriptsize $8(N_f-1)(N_f+1)(3N_f^4-2N_f^2+6)$  \\ \hline 
 \hline
\scriptsize SO($N_c$) & \scriptsize $ \begin{array}{l}
{{\rm SU}(N_f)}/{{\rm SO}(N_f)}\\
\quad\quad {\tiny (N_f\ge 3)} 
\end{array}$
 & \scriptsize $\frac12(N_f+2)(N_f-1)$ & \scriptsize $\frac{1}{12}N_f(N_f^2-1)(N_f^2-4)$ & \scriptsize $(N_f-1)(N_f+2)(3N_f^4+7N_f^3-2N_f^2-12N_f+24)$  \\ \hline 
 \hline
\scriptsize Sp($N_c$) &\scriptsize  $ \begin{array}{l}
{\rm SU}(2N_f)/{\rm Sp}(2N_f)\\
\quad\quad {\tiny (N_f\ge 2)} 
\end{array}$
 &\scriptsize $(2N_f+1)(N_f-1)$ & \scriptsize $\frac23N_f(N_f^2-1)(4N_f^2-1)$ &\scriptsize  $4(N_f-1)(2N_f+1)(6N_f^4-7N_f^3-N_f^2+3N_f+3)$   \\ \hline 
\end{tabular}  
\caption{Summary of group theory factors in the cases with nonzero WZW terms, quoted from Ref. \cite{Hochberg:2014kqa} and Ref. \cite{Witten:1983tx}. \label{Table:WZWsummary}} 
\end{center}
\end{table} 
 \end{widetext}

\section{Dark U(1) for SIMP dark mesons }\label{Sec:darkU(1)}

 The dark U(1)$_D$ charges of dark quarks are closely related to the form of quark mass terms.
 For SU($N_c$) gauge groups, quarks and anti-quarks belong to fundamental and anti-fundamental representations, respectively, {\it i.e.} complex representations, so Dirac mass terms such as $(m_q)_{ij}\overline{q}_iq_j$ are allowed. In this case, dark quarks can be vector-like under U(1)$_D$ so the model is automatically free from gauge anomalies. 
But, if U(1)$_D$ is unbroken,  dark mesons can be unstable in general, because dark mesons can decay fast into a pair of massless dark photons $\gamma_D$, in the presence of AVV chiral anomalies.
Even if the dark meson decays from AVV anomalies can be forbidden by appropriate U(1)$_D$ charge assignments, such as universal charges up to sign \cite{Bar:2001qk}, 
we cannot prohibit a dark meson self-annihilation in the form of $\pi\pi \to \pi \gamma_D$ through AAAV anomalies\footnote{Effects of both AVV and AAAV anomalies are encoded in the gauged WZW term \cite{Witten:1983tw,Kaymakcalan:1983qq}.} \cite{Bardeen:1969md}.
Then, for the $3\to2$ annihilation to be a dominant process for freeze-out, the dark gauge coupling for a unbroken $U(1)_D$ must be extremely small so the gauge kinetic mixing does not give an enough scattering cross section of dark mesons off the SM particles at freeze-out.
  
For our later discussion on $SU(N_c)$ confining groups, we take the U(1)$_D$ compatible with SIMP dark mesons to be spontaneously broken so that dark photon gets massive.
  For dark photon mass $m_V>m_\pi$, the $\pi\pi \to \pi \gamma_D$ processes from AAAV anomalies are kinematically forbidden. As will be discussed in the next section, in the presence of a gauge kinetic mixing between dark photon and the SM U(1)$_Y$, the off-shell processes, $\pi\pi \to \pi \gamma^*_D\rightarrow \pi e^- e^+$, opens up but it turns  out to be suppressed as compared to the $3\rightarrow 2$ processes.

   For SO($N_c$) and Sp($N_c$) gauge groups,  on the other hand, quarks in the fundamental representation, belong to real and pseudo-real representations, respectively, so there is no distinction between quarks and anti-quarks.
   As a result, only the Majorana mass terms are allowed. 
   Denoting Weyl spinor indices as $\alpha, \beta, \cdots$, gauge multiplet indices as $r, s, \cdots$, and flavor indices as $i, j, \cdots$, dark quark mass terms appear as 
   \dis{{m_q}^{(rs)(ij)}q^\alpha_{r,i}q_{\alpha s, j}+{\rm h.c.},} 
   in which ${m_q}^{(rs)(ij)}=m_q \delta^{rs}\delta^{ij}$ for SO($N_c$) and ${m_q}^{(rs)(ij)}=m_q J^{rs}J^{ij}$ for Sp($N_c$) gauge group, where $J\equiv i\sigma_2\otimes {\mathbb I}$ is an antisymmetric second rank tensor.
   In this case, dark quarks are only chiral under U(1)$_D$. Then, only after the U(1)$_D$ is broken spontaneously, dark quarks obtain masses  so does dark photon.

 When dark quarks are chiral under U(1)$_D$,  a special care is needed.
 Suppose that a chiral U(1)$_D$ is spontaneously broken and has a gauge kinetic mixing with the SM U(1)$_Y$, as for the case with a vector-like $U(1)_D$. 
   First of all, there should be no gauge anomalies, such as $G_c-G_c-$U(1)$_D$ or U(1)$_D$-U(1)$_D$-U(1)$_D$ anomalies. 
   Secondly, even for dark photon mass with $m_V>m_\pi/2$, the AVV anomalies could induce the decay of dark mesons into SM particles through the kinetic mixing, such as $\pi \to \gamma_D^* \gamma_D^* \to (e^+e^-)(e^+e^-)$. Therefore, the AVV anomaly terms should be forbidden.
   Finally, Partially Conserved Axial Current (PCAC) can couple to dark photon linearly such as $F\partial_\mu \pi V^\mu$ in general.
   This results in the decay of dark mesons into a pair of SM fermions, so it is dangerous as well.
   These challenges with chiral U(1)$_D$ can be overcome by considering appropriate U(1)$_D$ charge assignments, possibly calling for extra heavy dark quarks. In our work, however, we won't discuss this interesting case. 
   
   In order to make the discussion simple, we consider vector-like dark quarks under U(1)$_D$, that is spontaneously broken and has a kinetic mixing with SM U(1)$_Y$, and restrict ourselves to SU($N_c$) confining gauge symmetry. The physical results are not so different for the SO($N_c$) and Sp($N_c$) gauge groups that need the breakdown of a chiral U(1)$_D$ for nonzero quark masses, as far as heavy dark quarks do not have order one Yukawa couplings.   
   
 Dark mesons are pseudo-Goldstone bosons resulting from a spontaneous breaking of $G_f=$SU$(N_f)\times$SU($N_f$) down to $H$=SU($N_f$).
 As a minimal choice for a nonzero WZW term, we take $N_f=3$.
Furthermore, for the absence of the AVV anomalies, the U(1)$_D$ charge operator $Q_D$ must satisfy Tr$(Q_D^2 \lambda_a)=0$ with $\lambda_a$ being Gell-Mann matrices.
Thus, we choose the charge matrix to be
\be
Q_D=\left(\begin{array}{ccc} 1 & 0 & 0 \\ 0 & -1 & 0 \\ 0  & 0 & -1 \end{array}\right).  \label{charges}
\ee
  Then, dark mesons are written as an SU($3$)-valued matrix, $U(x)\equiv {\rm exp}(2i\sum_a \lambda_a \xi^a)$, where
  \begin{widetext}
    \dis{\sum_a \lambda_a \xi^a=\frac{\sqrt2}{F}\left[
\begin{array}{ccc}
\frac{1}{\sqrt2}\tilde{\pi}^0+\frac{1}{\sqrt6}\tilde{\eta}^0 & \tilde{\pi}^+ & \tilde{K}^+  \\
\tilde{\pi}^- & -\frac{1}{\sqrt2}\tilde{\pi}^0+\frac{1}{\sqrt6}\tilde{\eta}^0 & \tilde{K}^0  \\
\tilde{K}^- & \overline{\tilde{K}}^0 & -\sqrt{\frac{2}{3}}\tilde{\eta}^0 \\
\end{array}\right].}
  \end{widetext}
Due to the absence of AVV anomalies, neutral dark mesons are protected from the decays, $\tilde{\pi}^0, \tilde{\eta}^0 \to 2\gamma_D$.
One-loop corrections just rescale the vertices of AVV anomalies \cite{Donoghue:1988ct}. In our case, the stability of neutral mesons is guaranteed by charge assignments in Eq. (\ref{charges}) thanks to non-renormalization of AVV anomalies at all orders.  
Then,  the kinetic term $(F^2/16){\rm Tr}(D_\mu U D^\mu U^{-1})$, where the covariant derivative is $D_\mu U=\partial_\mu U+ig_D[Q_D, U]V_\mu$, with $g_D$ being the dark gauge coupling (or $\alpha_D\equiv g^2_D/4\pi$ being the dark structure constant), provides the leading interactions between dark mesons and dark photon $V_\mu$, 
   \dis{{\cal L}_{D \rm int}=&-i2g_D(\partial_\mu \tilde{K}^+ {\tilde K}^--\tilde{K}^+ \partial_\mu {\tilde K}^-+\partial_\mu \tilde{\pi}^+ \overline{\tilde \pi}^--\tilde{\pi}^+ \partial_\mu\overline{\tilde \pi}^-)V^\mu
   \\
&+4g_D^2(\tilde{K}^+\tilde{K}^-+\tilde{\pi}^+\tilde{\pi}^-)V_\mu V^\mu.}
 
 A remark on the effect of dark photon couplings on the mass splitting is in order. 
 The U(1)$_D$ charge assignment that we take makes some of dark mesons charged under U(1)$_D$, resulting in a dark meson mass splitting coming from $\alpha_D\Lambda^4{\rm Tr}(QUQU^{-1})$, that violates flavor symmetry explicitly.
 If the dark meson mass splitting is large enough, the only lightest dark meson remains eventually as a result of SU($N_c$) or U(1)$_D$ interactions.
 However, U(1)$_D$ mass contribution is small for a perturbatively small $\alpha_D$. 
 Since the SIMP scenario works for $m_\pi/F \gtrsim 4$ \cite{Hochberg:2014kqa}, for $\alpha_D =1/4\pi$, the dark photon contribution to the mass splitting is as small as $\Delta m_\pi^2\lesssim \alpha_D\Lambda^4/F^2 \sim \alpha_D F^2 \sim {\cal O}(10^{-2})m_\pi^2$, {\it i.e.} less than 10\%.
 Therefore, the dark meson mass degeneracy is a good approximation. Henceforth, we assume the U(1)$_D$ charges given in Eq.~(\ref{charges}).

\section{SIMP dark mesons with SU($N_c$) confining group }\label{Sec:kineticmix}

We consider a gauge kinetic mixing between the U(1)$_D$ gauge boson ($V_\mu$) and the U(1)$_Y$ gauge boson ($B_\mu$), given by
  \dis{{\cal L}_{\rm U(1)_D}=-\frac14 V_{\mu\nu}V^{\mu\nu}-\frac14 B_{\mu\nu} B^{\mu\nu}-\frac{\sin\chi}{2}V_{\mu\nu}B^{\mu\nu} +\frac12 m_V^2 V_\mu V^\mu.}
  After diagonalizing gauge kinetic and mass terms by  
      \dis{\left[
\begin{array}{c}
B_\mu \\ W^3_\mu \\ V_\mu
\end{array}\right]=
  \left[
\begin{array}{ccc}
c_W & -s_Wc_\zeta+t_\chi s_\zeta & -s_Ws_\zeta-t_\chi c_\zeta  \\
s_W & c_Wc_\zeta & c_W s_\zeta  \\
0 & -\frac{s_\zeta}{c_\chi} & \frac{c_\zeta}{c_\chi} \\
\end{array}\right]
\left[
\begin{array}{c}
A_\mu \\ Z_\mu \\ A'_\mu
\end{array}\right],}
where
\dis{\tan 2\zeta=\frac{m_Z^2s_W\sin2\chi}{m_V^2-m_Z^2(c_\chi^2-s_W^2s_\chi^2)},}
three mass eigenstates $(A_\mu,Z_\mu, A'_\mu)$ are interpreted as photon, $Z$-boson, and dark photon, respectively, with the masses of the latter two being 
\begin{widetext}
\dis{m_\pm^2=\frac12\Big[m_Z^2(1+s_W^2t_\chi^2)+\frac{m_V^2}{c_\chi^2}\pm\sqrt{\Big(m_Z^2(1+s_W^2t_\chi^2)+\frac{m_V^2}{c_\chi^2}\Big)^2-\frac{4}{c_\chi^2}m_Z^2m_V^2}\Big].}
\end{widetext}
We get  $m_+^2\simeq m_Z^2$ and $m_-^2\simeq m_V^2$ in the $\chi \to 0$ limit.
To these gauge bosons, electromagnetic (EM) current $J_{\rm EM}$, neutral $Z-$current $J_Z^\mu$, and dark sector current $J_D^\mu$ couple, as
\bea
{\cal L}_{\rm int}&=& A_\mu J_{\rm EM}^\mu +Z_\mu\Big[(c_W s_\zeta t_\chi)J_{\rm EM}^\mu+(c_\zeta-s_Wt_\chi s_\zeta)J_Z^\mu-\frac{s_\zeta}{c_\chi}J_D^\mu\Big]  \nonumber 
\\
&&+{A'}_\mu\Big[ (-c_W c_\zeta t_\chi) J_{\rm EM}^\mu+(s_\zeta+  s_Wt_\chi c_\zeta )J_Z^\mu+\frac{c_\zeta}{c_\chi}J_D^\mu\Big] .
\eea
The leading interaction between dark and SM sectors is the dark photon coupling to EM current with shifted charges by $(c_Wc_\zeta t_\chi)A'_\mu J_{\rm EM}^\mu$.
 The kinetic mixing parameter, $\epsilon_\gamma \equiv c_Wc_\zeta t_\chi$, and the dark photon mass, $m_V$, are constrained by various experiments.
 
 For the range $0.02\gev<m_V <10.2\gev$, the recent BaBar experiment shows that $\epsilon_\gamma \lesssim 6\times 10^{-4}$ from the observation of $e^+e^-\to \gamma \gamma_D*\to \gamma (\ell^+\ell^-)$ \cite{Lees:2014xha}.
 The LHC experiments provide bounds, $\epsilon_\gamma \lesssim 5\times 10^{-2}$ for $20\gev < m_V < 30\gev$, from the analysis of a new Higgs decay mode, $h \to Z \gamma_D$, based on CMS8 \cite{Curtin:2013fra}, as well as $\epsilon_\gamma \lesssim 10^{-2}$ for $30\gev < m_V < 70\gev$  and $m_V > 10^2 \gev$ from the Drell-Yan (DY) $\gamma_D$ production giving di-lepton signal \cite{Cline:2014dwa, Curtin:2014cca} based on Refs. \cite{Chatrchyan:2013tia}. 
 These constraints are, however, imposed under the assumption that dark photon decays into SM particles only \cite{Batell:2009yf}.
  In our case, dark photon can decays mainly into a pair of dark mesons, if kinematically allowed, {\it i.e.} $m_V>2m_\pi$, so the decay branching fraction into a pair of visible SM particles, ${\rm SM}_i$, is modified to be
  \dis{{\rm BR}(\gamma_D \to {\rm SM}_i)=\frac{\Gamma(\gamma_D \to  {\rm SM}_i)}{\sum_i\Gamma(\gamma_D\to {\rm SM}_i)+{\Gamma(\gamma_D \to \pi\pi)}}\simeq \Big(\frac{\epsilon_\gamma^2 \alpha}{\alpha_D}\Big)\frac{\Gamma(\gamma_D \to  {\rm SM}_i)}{\sum_i \Gamma(\gamma_D\to {\rm SM}_i)},}
  where $\Gamma(\gamma_D \to  {\rm SM}_i)/\sum_i\Gamma(\gamma_D\to {\rm SM}_i)$ in the last equality corresponds to the old branching ratio without invisible decays.
   Since the experimental limits on the visible modes depend on $\epsilon_\gamma^2 {\rm Br}(\gamma_D \to e^+e^-)$, the bound on $\epsilon_\gamma$ gets weaker by a factor $[\alpha_D/(\epsilon_\gamma^2\alpha)]^{1/2}$.
 Moreover, electroweak precision test (EWPT) provides stringent bounds, $\epsilon_\gamma \lesssim 2\times 10^{-2}$ for $10\gev < m_V < 80\gev$, and $\epsilon_\gamma \lesssim 2.5\times 10^{-3}$ for $m_V \simeq m_Z$, 
 which is not affected by the invisible decays discussed above \cite{Curtin:2014cca}.    
 
 Dark photon also couples to $Z-$current.
 For $m_V \ll m_Z$, however, the  $Z-$current coupling does not give a significant contribution, since the mixing angle approximated by $\zeta \simeq -s_W \chi$ makes the coefficient for dark photon coupling to $Z-$current, $\epsilon_Z \equiv s_\zeta+s_W t_\chi c_\zeta$, vanish at the leading order.
 On the other hand, for the dark photon mass being around the $Z-$boson mass, the mixing angle gets larger as $\zeta \simeq (m_Z^2 t_W\epsilon_\gamma)/(m_V^2-m_Z^2)$, so we cannot ignore it any longer and the U(1)$_D$ gauge boson is interpreted as a `dark $Z$ boson' \cite{Davoudiasl:2012ag}.
 This makes the lower bound on $\epsilon_\gamma$ less stringent.

 We are now in a position to consider the conditions on $m_V$ and $\epsilon_\gamma$ that are consistent with the SIMP mechanism.
 At freeze-out temperature, $T_F \simeq m_\pi/20 \sim (5-50)$MeV, photon, electron/positron, and neutrinos are relativistic particles in thermal equilibrium, and muon and pion begin to be non-relativistic. 
 At that moment, the $3\to2$ self-annihilation from the WZW term is dominant over the other possible annihilation processes whereas dark meson-SM particle scattering processes do not decouple yet, provided that 
 \dis{ n_{\rm DM}\langle \sigma v \rangle_{\rm ann} < n_{\rm DM}^2\langle \sigma v^2 \rangle_{3\to2} <n_{\rm SM}\langle \sigma v \rangle_{\rm scatt},}
 at the freeze-out temperature,
 where the number density of a bosonic or fermionic SM particle is given by
 \dis{n_{\rm SM}=\frac{g}{2\pi^2}T_F^3 \int_0^\infty \frac{x^2}{e^{\sqrt{x^2+(m_{\rm SM}/T_F)^2}}\mp 1}.\nonumber} 
 
 In order to prevent a pair of dark mesons from annihilating into a pair of dark photons due to the gauge interactions with $\pi\pi A_\mu A^\mu$,  we require $m_V>m_\pi$.
 In this region, the $\pi \pi \to {\rm SM}\, {\rm SM}$ annihilation rate, estimated as $n_{\rm DM}\times[{\cal O}(10^2)\alpha \alpha_D \epsilon_\gamma^2 m_\pi^2/(N_\pi m_V^4)]$, is smaller than the $3\to2$ annihilation rate, if
 \dis{\epsilon_\gamma \lesssim  0.05 \Big(\frac{N_c}{10}\Big)\Big(\frac{m_V}{10\gev}\Big)\Big(\frac{0.5\gev}{m_\pi}\Big)^{5/2}.}
 For typical parameters such as $T_F \simeq m_\pi/20$, $F\simeq m_\pi/5$ and $\alpha_D \simeq 1/4\pi$, the above condition is fulfilled for any value of $\epsilon_\gamma$  satisfying the upper bounds given by ground-based experiments.
 The only exception appears around $m_V = 2m_\pi$, where $\pi \pi \to {\rm SM}\, {\rm SM}$ annihilation rate is improved due to a resonance from $1/(4m_\pi^2-m_V)^2$.
  
 As for the dark meson scattering off the SM particles, the dominant process is $\pi+e^\pm \to \pi +e^\pm$ through the $t-$channel process, whereas $\pi +\gamma \to \pi +\gamma$ is suppressed by a   double kinetic mixing. 
 Ignoring the lepton masses, we obtain the scattering cross section for $\pi+\ell \to \pi + \ell$ averaged over the number of dark mesons $N_\pi$ as 
 \begin{widetext}
 \dis{\langle \sigma v \rangle_{{\rm scatt},\ell}&= \frac{4}{N_\pi}\cdot \Big[192\epsilon_\gamma^2+\frac{24(8s_W^4-4s_W^2+1)\epsilon_Z^2 }{c_W^2s_W^2}\Big] \pi\alpha \alpha_D\, \frac{m^2_\pi}{m^4_V}\left(\frac{T_F}{m_\pi}\right),}
 \end{widetext}  
  where a factor 4 represents the degrees of freedom of U(1)$_D$ charged mesons, $K^\pm$ and $\pi^\pm$.
  On the other hand, the scattering cross sections of dark mesons off the neutrinos, $\pi +\nu \to \pi+\nu$ and  the SM pion, $\pi+ \pi_{\rm SM}^\pm \to \pi + \pi_{\rm SM}^\pm$, are given by 
  \begin{widetext}
   \dis{
 &\langle  \sigma v\rangle_{{\rm scatt},\nu}
= \frac{4}{N_\pi}\cdot \frac{24\pi\alpha\alpha_D\epsilon_Z^2 m^2_\pi }{c_W^2s_W^2 m_V^4}\,\left(\frac{T_F}{m_\pi}\right),
\\   
&   \langle \sigma v\rangle_{{\rm scatt},\pi}=\frac{4}{N_\pi}\cdot\Big[192\epsilon_\gamma^2+\frac{192\epsilon_Z^2}{c_W^2 s_W^2}\frac14 (1-2s_W^2)^2\Big] \pi\alpha \alpha_D\, \frac{m^2_\pi}{m^4_V} \left(\frac{T_F}{m_\pi}\right).}
 From the condition,
 \dis{n_{\rm DM}^2\langle \sigma v^2 \rangle_{3\to2} <n_{\rm SM}\langle \sigma v \rangle_{\rm scatt}=\sum_{\ell=e,\mu} n_{\ell} \langle \sigma v \rangle_{{\rm scatt},\ell} +n_\nu \langle \sigma v \rangle_{{\rm scatt},\nu}+  n_{\pi_{\rm SM}} \langle \sigma v\rangle_{{\rm scatt},\pi}, }
  \end{widetext}
the larger $m_V$, the stronger the lower bound on $\epsilon_\gamma$ gets, according to
  \dis{\alpha_D \epsilon_\gamma \Big(\frac{m_\pi}{m_V}\Big)^2 \gtrsim 10^{-8}.\label{Eq:lowerbound}}

  \begin{widetext}
\begin{figure}[t]
 \begin{center}
   \includegraphics[width=8cm]{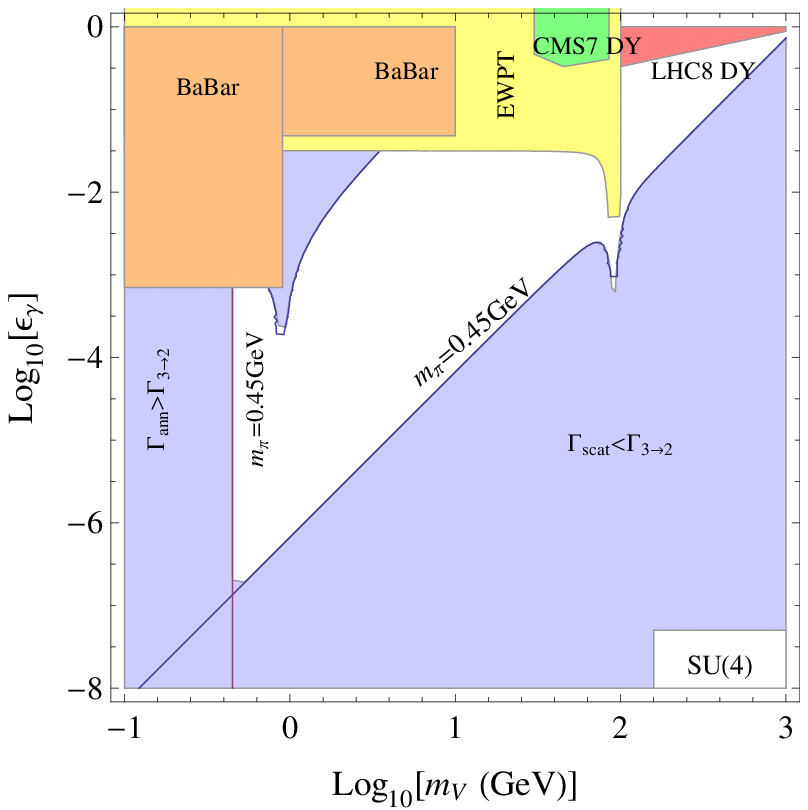}
  \includegraphics[width=8cm]{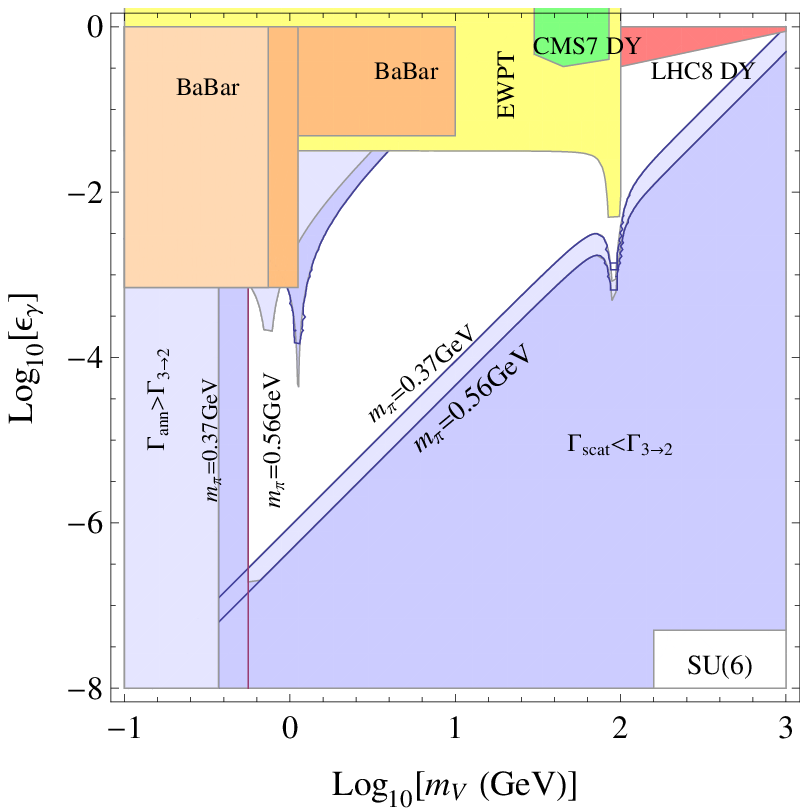}
  \includegraphics[width=8cm]{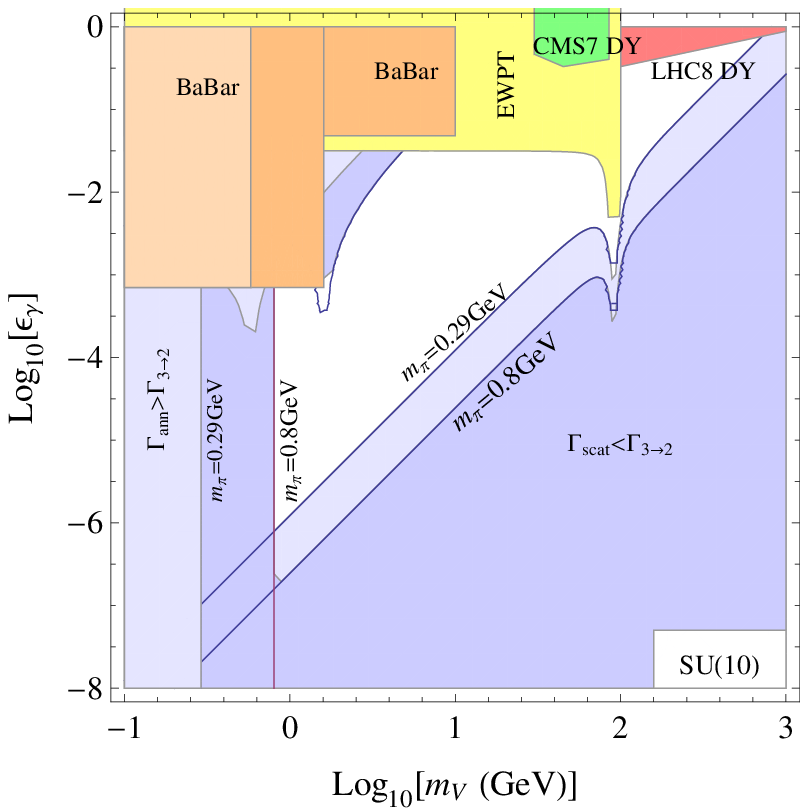}
  \vspace{-1em}
 \end{center}
 \caption{ Bounds on $m_V$ vs $\epsilon_\gamma$ for dark mesons being compatible with the SIMP scenario for $1/4\pi$. 
 Three figures correspond to $G_c=$SU$(4)$, SU$(6)$, and SU$(10)$, respectively. Imposed constraints, distinguishable by colors, are written explicitly, while the allowed parameter space is uncolored.   For $m_V>2m_\pi$, BaBar and LHC bounds are rescaled taking $\gamma_D \to 2\pi$ invisible decay into account.}
 \label{fig:SIMPU(1)}
\end{figure}
\end{widetext}

  \begin{widetext}
\begin{figure}[t]
 \begin{center}
   \includegraphics[width=8cm]{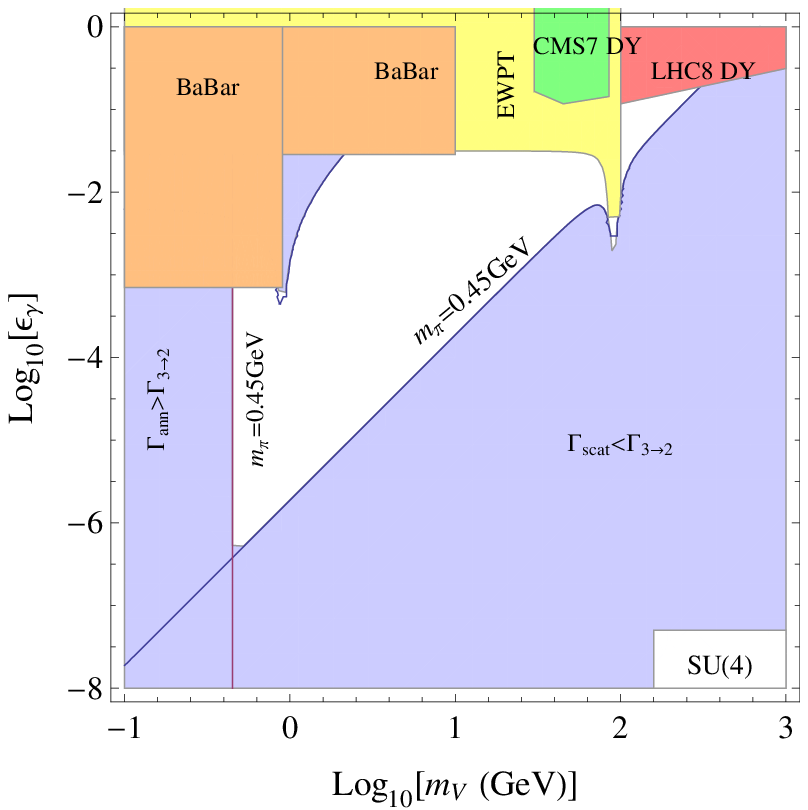}
  \includegraphics[width=8cm]{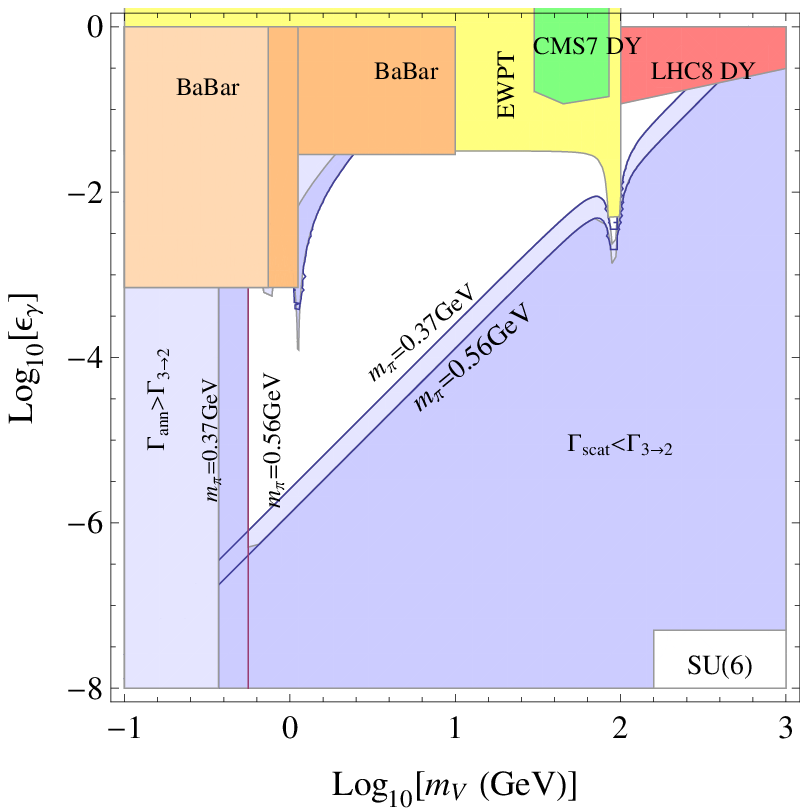}
  \includegraphics[width=8cm]{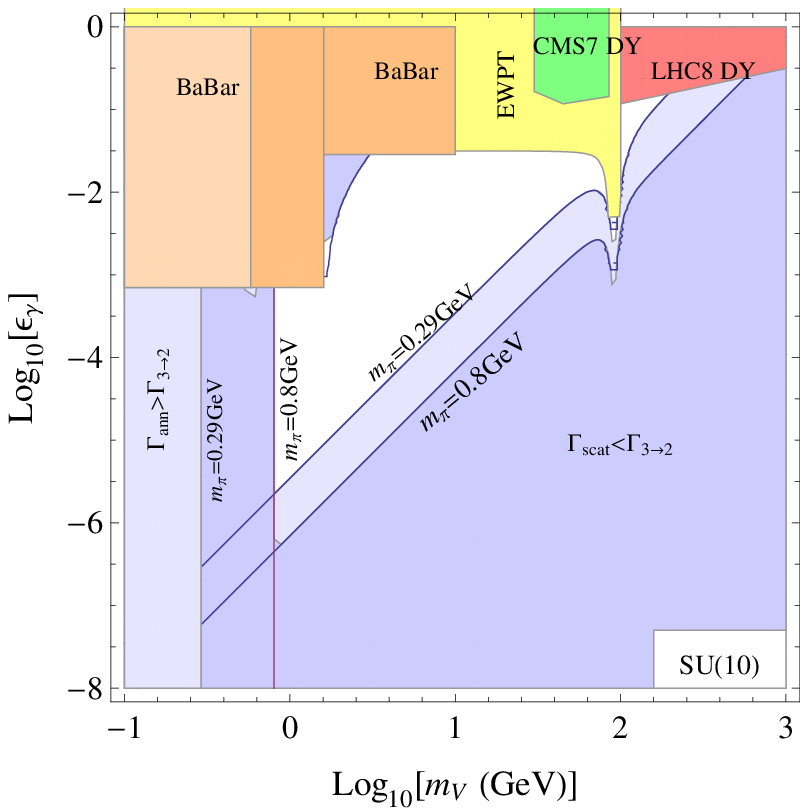}
  \vspace{-1em}
 \end{center}
 \caption{ Bounds on $m_V$ vs $\epsilon_\gamma$ for dark mesons being compatible with the SIMP scenario for $\alpha_D=0.01$.}
 \label{fig:SIMPU(1)2}
\end{figure}
\end{widetext}

 We present constraints on $m_V$ and $\epsilon_\gamma$ for several cases with different confining groups in Fig. \ref{fig:SIMPU(1)} and Fig. \ref{fig:SIMPU(1)2}, where $\alpha_D=1/4\pi$ and $\alpha_D=0.01$ were taken, respectively,
   and that dark meson masses are assumed to be degenerate.
  For $N_c=4$, the minimal $N_c$ that the SIMP mechanism works, only $m_\pi\simeq 0.45\gev$ is allowed because the perturbativity condition, $x \equiv m_\pi/F< 2\pi$, and the self-interaction bound, $\sigma_{\rm self}/m_\pi \lesssim 1 {\rm cm}^2/{\rm g}$, almost coincide \cite{Hochberg:2014kqa}.
  For $N_c=6$ and $N_c=10$, a wider range of dark meson masses are allowed such as $0.37\gev<m_\pi <0.56\gev$ and $0.26\gev < m_\pi < 0.8\gev$, respectively, and $x$ is fixed by the DM relic density. 
  In both cases, the upper bounds satisfy $x=2\pi$ and the lower bounds satisfy $x=5.48$ and $4.6$, respectively.
  
We note that dark photon is taken to be heavier than dark mesons in order for the $3\to2$ annihilation to dominate over the $\pi\pi \to \gamma_D \gamma_D$ annihilation. As a result, there appears a lower bound, $\epsilon_\gamma \gtrsim 10^{-7}$ at $m_V=m_\pi$, due to Eq.~(\ref{Eq:lowerbound}).
On the other hand,  the AAAV anomalies induce the annihilation process such as $\pi(k_1)\pi(k_2)\to \pi(k_3)e^-(p_1)e^+(p_2)$ through off-shell dark photon, where momenta of particles in the process are explicitly written.  The interaction vertex contains $\epsilon_{\mu\nu\rho\sigma}k_1^\nu k_2^\nu k_3^\rho \overline{v}(p_2)\gamma^\sigma u(p_1)$ term, which vanishes in non-relativistic limit, $k_1 \simeq k_2 \simeq (m_\pi, \vec{0})$. Thus, the annihilation cross section for this process is estimated as $\alpha\alpha_D \epsilon_\gamma^2 m_\pi^6 T_F^2/(N_\pi m_V^4 F^6)$. 
Therefore, as compared to the annihilation cross section of $\pi\pi\to e^-e^+$, which is smaller than the one for the $3\to2$ processes, the annihilation cross section of $\pi\pi\to \pi e^-e^+$ is suppressed by $m_\pi^4 T_F^2/[(2\pi)^3 F^6] \simeq m_\pi^6/(2\pi F)^6 $, within the valid regime of chiral perturbation theory.

The lower bound on $\epsilon_\gamma$ for a given $m_V$ follows from the estimation given by Eq. (\ref{Eq:lowerbound}), and the upper bound comes from ground-based experiments.
  We also find from Eq. (\ref{Eq:lowerbound}) that the SIMP condition requires a large kinetic mixing for a large dark photon mass, eventually constrained by ground-based experiments.
 In summary, for $N_c <10$,  dark photon masses are allowed up to $m_V\sim 10^3$ GeV with varying limits on $\epsilon_\gamma$.
 We note that there are more allowed values of $\epsilon_\gamma$ around $m_V \simeq m_Z$ due to the non-negligible contribution from $\epsilon_Z$. The bounds get stronger near $m_V = 2m_\pi$, where the dark meson annihilation into a pair of SM particles becomes enhanced as discussed previously.
   
\section{Conclusion}\label{Sec:Conclusion}

 We have considered an extension of the models with SIMP dark mesons by including a dark local U(1)$_D$ symmetry under which dark quarks are vector-like. 
 Dark mesons are still good candidates for SIMP DM,  as the chiral anomalies associated with U(1)$_D $ are absent.  In the presence of a gauge kinetic mixing between U(1)$_D$ and the SM U(1)$_Y$, the dark sector is communicated with  the SM particles through the $Z'$  portal so that it can be kept in kinetic equilibrium with the SM sector until the freeze-out in the SIMP scenario.
 
The SIMP conditions restrict the parameter space for dark photon mass $m_V$ and kinetic mixing $\epsilon_\gamma$, that is otherwise unconstrained by ground-based experiments.
Focusing on dark mesons living on the SU$(N_f)\times$SU$(N_f)/$SU$(N_f)$ flavor symmetry and taking the SU($N_c$) confining group for them,  we showed that the combination of  the SIMP conditions with various ground-based experiments searching for dark photon can restrict the parameter space to $m_\pi <m_V\lesssim 10^3\gev$ and $10^{-7} <\epsilon_\gamma< (10^{-3}-10^{-2})$, for dark gauge coupling of order one and $N_c <10$.

\section*{Acknowledgments}
We would like to thank Su Min Choi, Ji-Haeng Huh, Kwang Sik Jeong, Hye-Sung Lee and Jong-Chul Park for discussions.
The work of HML is supported in part by Basic Science Research Program through the National Research Foundation of Korea (NRF) funded by the Ministry of Education, Science and Technology (2013R1A1A2007919).  
 The work of MS is supported by IBS-R018-D1.






\end{document}